# Efficient Network for Non-Binary QC-LDPC Decoder

Chuan Zhang and Keshab K. Parhi
Department of Electrical and Computer Engineering
University of Minnesota, Twin Cities
Minneapolis, MN 55455, USA
{zhan0884, parhi}@umn.edu

*Abstract*—This paper presents approaches to develop efficient network for non-binary quasi-cyclic LDPC (QC-LDPC) decoders. By exploiting the intrinsic shifting and symmetry properties of the check matrices, significant reduction of memory size and routing complexity can be achieved. Two different efficient network architectures for Class-I and Class-II non-binary QC-LDPC decoders have been proposed, respectively. Comparison results have shown that for the code of the 64-ary (1260, 630) rate-0.5 Class-I code, the proposed scheme can save more than 70.6% hardware required by shuffle network than the state-of-the-art designs. The proposed decoder example for the 32-ary (992, 496) rate-0.5 Class-II code can achieve a 93.8% shuffle network reduction compared with the conventional ones. Meanwhile, based on the similarity of Class-I and Class-II codes, similar shuffle network is further developed to incorporate both classes of codes at a very low cost.

## I. INTRODUCTION

Compared with binary ones, LDPC codes defined upon Galois field $GF(q)$ with order higher than two have even more excellent error correction capabilities with proper encoding approach and code length [1]. However, with the improvement of decoding performance, higher computation complexity will follow. To this end, [2] proposed several sub-optimal selecting algorithms based on $n$-norm $\|\ \|_n$ construction. As $n$ decreases to 1, the optimal algorithm reduces to the Min-Max algorithm, which proves very suitable for practical purposes by achieving a good compromise between hardware costs and decoding performance. Meanwhile, a special class of non-binary LDPC codes named non-binary QC-LDPC codes are constructed with the architecture-aware scheme [3]. Even though, the few implementations of non-binary QC-LDPC decoders employing Min-Max algorithm still suffer from a high hardware cost [4]-[5]. Without exploiting the inherent geometry properties of QC-LDPC codes, those designs employ either a conventional bi-directional network or two shuffle networks for re-/shuffling, leading to a $q$-time increase of the network complexity.

In this paper, to make full use of benefits introduced by architecture-aware scheme, special emphasis has been placed on investigating the geometry properties of the corresponding **H** matrices. Rather than reconfiguring the global shuffle networks for each layer, the proposed approach employs two kinds of *local shuffle network* to eliminate the unnecessary network costs for Class-I and Class-II QC-LDPC codes, respectively. The designs are reconfigurable, memory efficient, highly parallel, and of low routing complexity. In order to demonstrate the advantages, both the 64-ary (1260, 630) rate-0.5 Class-I code and 32-ary (992, 496) rate-0.5 Class-II code are employed as examples. It is shown that, if flexibility is taken into account, 70.6% shuffle network cost can be reduced compared with the state-of-the-art designs for Class-I code and 93.8% for Class-II code. On the other hand, if flexibility is not a necessity, more memory and control logic can be eliminated. Moreover, a new *local shuffle network* which is compatible with both classes has been further proposed along with a minimum cost.

The remainder of this paper is organized as follows. In Section II, construction methods for both Class-I and Class-II codes are briefly reviewed. In Section III, geometry properties of both codes are investigated and summarized, respectively. Different layer partition choices have been proposed in Section IV. Section V describes the shuffle networks. The hardware costs estimation and comparisons with other designs are given in Section VI. Finally, Section VII concludes the whole paper.

## II. NON-BINARY QC-LDPC CODES CONSTRUCTION

Like their binary counterparts, non-binary QC-LDPC codes are initiated by the motivation of architecture-aware design [3]. Using two similar *array dispersions* of matrices, constructions for two classes of QC-LDPC codes, referred as Class-I and Class-II codes, are proposed as well. It is know that elements of $GF(q)$ can be represented in the power of primitive element $\alpha$: $\alpha^{-\infty} = 0$, $\alpha^0 = 1$, $\alpha^1$, ..., $\alpha^{q-2}$. The location vector $\mathbf{z}(\alpha^i)$ is defined as $\mathbf{z}(\alpha^i) = (z_0, z_1, \ldots, z_{q-2})$, where the $i$-th component $z_i = \alpha^i$, and all others are zeros. In addition, $\mathbf{z}(0)$ is the all-zero $(q-1)$-tuple. The definition of *circulant permutation matrix* (CPM) of $\mathbf{z}(\delta)$ is given by $(\mathbf{z}(\delta), \mathbf{z}(\alpha\delta), \ldots, \mathbf{z}(\alpha^{q-2}\delta))^T$ accordingly, where $\delta$ can be any element in $GF(q)$. Therefore, the construction steps for Class-I codes can be stated as follows:

---

**Construction for Class-I Non-Binary QC-LDPC Codes**

**1: Factorization:** $q - 1 = c \times n, \gcd(c, n) = 1$;

**2: Element definition:** $\beta = \alpha^c, \delta = \alpha^n$;

**3: Subgroup expansion:**
$$\begin{cases} G_1 = \{\beta^0 = 1, \beta, \ldots, \beta^{n-1}\}, \\ G_2 = \{\delta^0 = 1, \delta, \ldots, \delta^{c-1}\}; \end{cases}$$

**4: Matrix formation:**
$$\begin{cases} \mathbf{W}^{(1)} = [\mathbf{W}^{(1)}_{i,j}]_{0 \leq i < c, 0 \leq j < c}, \\ \mathbf{W}^{(1)}_{i,j} = [\delta^{j-i}\beta^k - \beta^l]_{0 \leq k < n, 0 \leq l < n}; \end{cases}$$

**5: Substitution:** Replace each entry of $\mathbf{W}^{(1)}$ by its CPM to get $\mathbf{A}^{(1)} = [\mathbf{A}^{(1)}_{i,j}]_{0 \leq i, j < q}$;

**6: Truncation:** $\mathbf{H}^{(1)} = \mathbf{A}^{(1)}(\gamma, \rho) = [\mathbf{A}^{(1)}_{i,j}]_{0 \leq i < \gamma, 0 \leq j < \rho}$;

**7: Output:** $\mathbf{H}^{(1)}$.

---

By simply changing the multiplications with additions, we obtain the construction steps for Class-II codes as follows. The similar construction steps for both codes yields resemblances in corresponding geometry properties and network designs, which are stated in Section III and V, respectively.

**Construction for Class-II Non-Binary QC-LDPC Codes**

**1: Factorization:** $q = 2^m$, $c = 2^{m-t}$, and $n = 2^t$;

**2: Elements definition:** $\begin{cases} f'_t = \{\alpha^0, \alpha^1, \ldots, \alpha^{t-1}\}, \\ f''_{m-t} = \{\alpha^t, \alpha^{t+1}, \ldots, \alpha^{m-1}\}; \end{cases}$

**3: Subgroup expansion:** $\begin{cases} \mathcal{F}'_t = \{0, \beta_1, \ldots, \beta_{2^t-1}\}, \\ \mathcal{F}''_{m-t} = \{0, \delta_1, \ldots, \delta_{2^{m-t}-1}\}; \end{cases}$

**4: Matrix formation:** $\begin{cases} \mathbf{W}^{(2)} = [\mathbf{W}^{(2)}_{i,j}]_{0 \le i,j < c}, \\ \mathbf{W}^{(2)}_{i,j} = [(\delta_i - \delta_j) + (\beta_k - \beta_l)]_{0 \le k,l < n}; \end{cases}$

**5: Substitution:** Replace each entry of $\mathbf{W}^{(2)}$ by its CPM to get $\mathbf{A}^{(2)} = [\mathbf{A}^{(2)}_{i,j}]_{0 \le i,j < q}$;

**6: Truncation:** $\mathbf{H}^{(2)} = \mathbf{A}^{(2)}(\gamma, \rho) = [\mathbf{A}^{(2)}_{i,j}]_{0 \le i < \gamma, 0 \le j < \rho}$;

**7: Output:** $\mathbf{H}^{(2)}$.

### III. PROPERTIES OF NON-BINARY QC-LDPC CODES

Although some apparent properties of non-binary QC-LDPC codes have been addressed by previous literatures [4]-[5], more geometry properties hidden behind the algebra architectures need to be revealed for efficient network designs.

#### A. Shifting Properties of Class-I Codes

According to the construction steps, one can verify the identity of $\mathbf{W}^{(1)}_{i,j}$ and its upper-left neighbor $\mathbf{W}^{(1)}_{(i-1) \bmod c, (j-1) \bmod c}$:

$$\begin{aligned} \mathbf{W}^{(1)}_{i,j} &= [\delta^{j-i} \beta^k - \beta^l] \\ &= [\delta^{((j-1) \bmod c) - ((i-1) \bmod c)} \beta^k - \beta^l] \\ &= \mathbf{W}^{(1)}_{(i-1) \bmod c, (j-1) \bmod c}. \end{aligned} \quad (1)$$

Similar permutation property holds at the lower level:

$$\begin{aligned} \mathbf{w}^{(1)}_{(i,j)(k,l)} &= \delta^{j-i} \beta^k - \beta^l \\ &= \beta(\delta^{j-i} \beta^{(k-1) \bmod n} - \beta^{(l-1) \bmod n}) \\ &= \beta \mathbf{w}^{(1)}_{(i,j)((k-1) \bmod n, (l-1) \bmod n)}. \end{aligned} \quad (2)$$

Moreover, with the definition of CPM, the useful properties of Class-I codes can be summarized as follows:

**Proposition 1** The Class-I non-binary QC-LDPC codes satisfy shifting properties at three different levels:

1) The $i$-th row of the base matrix $\mathbf{W}^{(1)}$ is exactly the 1-step right cyclic-shift of the $[(i-1) \bmod c]$-th row. Therefore, $\mathbf{W}^{(1)}_{i,j} = \mathbf{W}^{(1)}_{(i-1) \bmod c, (j-1) \bmod c}$;
2) The $k$-th row of the sub-matrix $\mathbf{W}^{(1)}_{i,j}$ is exactly the 1-step right cyclic-shift of the $[(k-1) \bmod n]$-th row multiplied by $\beta$, that is, $\mathbf{w}^{(1)}_{(i,j)(k,l)} = \beta \mathbf{w}^{(1)}_{(i,j)(k-1,l-1)}$;
3) The $m$-th row of the CPM corresponding to $\mathbf{w}^{(1)}_{(i,j)(k,l)}$ is the 1-step right cyclic-shift of the $[(m-1) \bmod n]$-th row multiplied by $\alpha$, which is given by the definition of CPM.

#### B. Symmetry Properties of Class-II Codes

Unlike Class-I codes, it is not that trivial to uncover the geometry properties of Class-II codes. Since the surjective function mapping from elements of subgroups $\mathcal{F}'_t$ and $\mathcal{F}''_{m-t}$ to power forms of $\alpha$ is not specified, candidate(s) for value assignment scheme is not unique. Given this degree of freedom, a specific surjective function which yields symmetry properties is introduced purposely. Without loss of generality, details of the surjective function are described with the subgroup $\mathcal{F}'_t$:

**Index Assignment of Surjective Function**

**1:** Suppose $\beta_i = \sum_{m=m_0}^{m_p} \alpha^m$, $\beta_j = \sum_{n=n_0}^{n_q} \alpha^n$, with $0 \le m, n < t$;

**2: if** $p < q$ **then** $i < j$;

**3: elseif** $p > q$ **then** $i > j$;

**4: else**

**5:**    **for** $l = 0, l++, l \le p$ **do**

**6:**        **if** $m_l < n_l$ **then** $i < j$ **break**;

**7:**        **else** $m_l > n_l$ **then** $i > j$ **break**;

**8:**    **endfor**

**9: endif**

According to the proposed function, it can be verified that for any element $\beta_i$ in sub-group $\mathcal{F}'_t$, Eq. (3) holds,

$$\beta_i + \beta_{(2^t-1)-i} = \beta_{2^t-1} \quad (3)$$

For sub-group $\mathcal{F}''_{m-t}$, similar symmetry property holds as well,

$$\delta_i + \delta_{2^{m-t}-1-i} = \delta_{2^{m-t}-1}. \quad (4)$$

Substituting Eq. (3) into the construction steps, we show that matrix $\mathbf{W}^{(2)}$ has the symmetry property shown in Eq. (5). Denote the sub-matrix of $\mathbf{W}^{(2)}$ by $\mathbf{W}^{(2)}_{i,j}$, then we have,

$$\mathbf{W}^{(2)}_{i,j} = \mathbf{W}^{(2)}_{c-j-1, c-i-1}, \quad (5)$$

Combining this fact with the matrix structure, it follows that $\mathbf{W}^{(2)}_{i,j}$ and its mirror about the anti-diagonal are identical. Similarly, Eq. (4) corresponds to the self-symmetry of $\mathbf{W}^{(2)}_{i,j}$ about its own anti-diagonal,

$$\mathbf{w}^{(2)}_{(i,j)(k,l)} = \mathbf{w}^{(2)}_{(i,j)(n-l-1, n-k-1)}. \quad (6)$$

On the other hand, for Class-II codes, both the base matrix $\mathbf{W}^{(2)}$ and its sub-matrix $\mathbf{W}^{(2)}_{i,j}$ are self-symmetric about their own diagonals. This is apparent from the 4$^{th}$ step of the construction for Class-II codes as follows,

$$\begin{aligned} \mathbf{W}^{(2)}_{i,j} &= [(\delta_i - \delta_j) + (\beta_k - \beta_l)] \\ &= [(\delta_j - \delta_i) + (\beta_k - \beta_l)] \\ &= \mathbf{W}^{(2)}_{j,i}, \end{aligned} \quad (7)$$

$$\begin{aligned} \mathbf{w}^{(2)}_{(i,j)(k,l)} &= (\delta_i - \delta_j) + (\beta_k - \beta_l) \\ &= (\delta_i - \delta_j) + (\beta_l - \beta_k) \\ &= \mathbf{w}^{(2)}_{(i,j)(l,k)}. \end{aligned} \quad (8)$$

The above properties help to derive the following proposition:

**Proposition 2** The Class-II non-binary QC-LDPC codes satisfy the geometry properties at three different levels:

1. The base matrix $\mathbf{W}^{(2)}$ is symmetric about its diagonal and anti-diagonal, i.e., $\mathbf{W}^{(2)}_{i,j} = \mathbf{W}^{(2)}_{j,i}$ and $\mathbf{W}^{(2)}_{i,j} = \mathbf{W}^{(2)}_{c-j-1, c-i-1}$;
2. The sub-matrix $\mathbf{W}^{(2)}_{i,j}$ is also symmetric about its diagonal and anti-diagonal, i.e., we have $\mathbf{w}^{(2)}_{(i,j)(k,l)} = \mathbf{w}^{(2)}_{(i,j)(n-l-1, n-k-1)}$ and $\mathbf{w}^{(2)}_{(i,j)(k,l)} = \mathbf{w}^{(2)}_{(i,j)(l,k)}$;
3. Each row of one CPM $\mathbf{w}^{(2)}_{(i,j)(k,l)}$ is the right cyclic-shift of the row above it multiplied by $\alpha$ and the first row is the right cyclic-shift of the last row multiplied by $\alpha$.

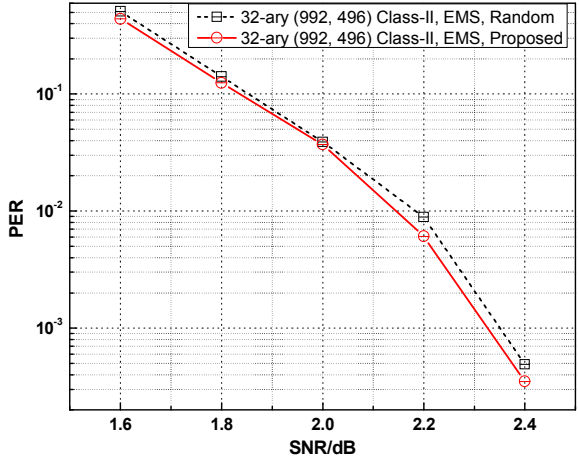

Figure 1: Performance comparison between two surjective functions.

Choose $m = 5$, and $t = 2$, we construct a 32-ary (992, 496) rate-0.5 Class-II code. Another Class-II code constructed with random surjective function is used as a benchmark. Decoding performances with EMS algorithm and maximum 10 iterations are illustrated in Fig. 1. It is shown that the performance of Class-II code with the proposed method is as good as that of the random one. Therefore, the proposed approach introduces symmetry properties without affecting the decoding advantage.

## IV. LAYER PARTITION CHOICES OF QC-LDPC CODES

In order to reduce the number of decoding iterations and make best use of the geometry properties, the layered decoding schedule is incorporated with the Min-Max algorithm here. The $k$-th iteration for layer $t$ can be formulated as follows:

**Layered Decoding for Min-Max Algorithm**

1: $L_{cv}^{k,t}(a) = L_v^{k,(t-1)}(a) - R_{cv}^{(k-1),t}(a)$;

2: $R_{cv}^{k,t}(a) = \min_{\substack{(a_{v'})_{v' \in \mathcal{H}(c) \setminus (v)} \\ \in \mathcal{L}(c|a_v=a)}} (\max_{v' \in \mathcal{H}(c) \setminus (v)} L_{cv'}^{k,t}(a_{v'}))$;

3: $L_v^{k,t}(a) = L_{cv}^{k,(t-1)}(a) + R_{cv}^{k,t}(a)$.

Stated in **Proposition 1** and **2**, both classes of non-binary QC-LDPC codes have a nice algebraic construction which can be easily accommodated with the layered decoding scheme. Along with the constraint of at most 1 column weight in each layer, two layer partition options can be proposed as follows:

1. Choose each sub-block row of $\mathbf{W}_{i,j}^{(1)}$ or $\mathbf{W}_{i,j}^{(2)}$ as one layer, which consists of $(q-1)$ rows. This option is defined as the *Layer-I* choice;
2. Choose each row of CPM $\mathbf{w}_{(i,j)(k,l)}^{(1)}$ or $\mathbf{w}_{(i,j)(k,l)}^{(2)}$ as one layer, which consists of only one row. This option is defined as the *Layer-II* choice.

## V. LOCAL SHUFFLE NETWORKS FOR BOTH CODES

The architecture of the $(u, v)$ non-binary QC-LDPC decoder is shown in Fig. 2. $u = \rho(q-1)$ is the code length, $u-v = \gamma(q-1)$ is the number of check bits, $w$ is the layer height. It is composed of $w$ CNUs, a de-/permutation block, a *global shuffle network* (GSN), and $u$ VNUs with a *local shuffle network* ($\Pi$). In what follows, with proposed geometry properties, different reduced-complexity shuffle networks for both codes are presented.

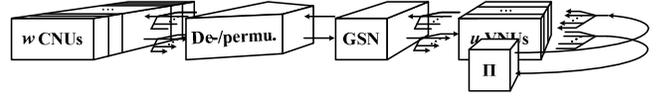

Figure 2: Block diagram of the layered non-binary QC-LDPC decoder.

### A. Local Shuffle Network for Class-I Codes

#### 1) Generating Algorithm for Local Shuffle Network

Apparently, **Proposition 1.2** and **1.3** only differ in the value of the multiplicand ($\beta$ or $\alpha$). Without loss of generality, the *Layer-I* decoding scheme is chosen as an example.

**Scheduling Algorithm for Local Shuffle Network - I**

1: for all $0 \leq i < \rho$ do
2:     for all $0 \leq j < q-1$ do
3:         Pass the *extrinsic* result of last layer from
4:         $\text{VNU}_{i(q-1)+j}$ to $\text{VNU}_{[(i-1) \bmod n](q-1)+(j-c) \bmod (q-1)}$
5:     endfor
6: endfor

The index of each VNU can be rewritten in the form of $i(q-1)+j$ (for short $(i, j)$). For the example depicted in Fig. 3, the index of $\text{VNU}_7$ can be rewritten as $(2, 1)$. Based on the new scheduling algorithm, the destination index is $(1, 0)$. Therefore, the *extrinsic* message should be transferred from $\text{VNU}_7$ to $\text{VNU}_3$ ($1 \times 3 + 0 = 3$), which matches the previous analysis.

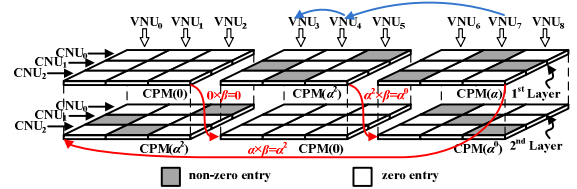

Figure 3: Layered decoding example of the 4-ary (9, 3) rate-⅓ Class-I code.

#### 2) Achitecture of Local Shuffle Network

It is observed that the inter-layer shuffle scheduling is irrelevant of the current layer index. That is, no matter what number $i$ is, the *extrinsic* message transfering between the $i$-th layer and the $(i+1)$-th layer is exactly the same, which can be implemented by using fixed interconnections shown in Fig. 4.

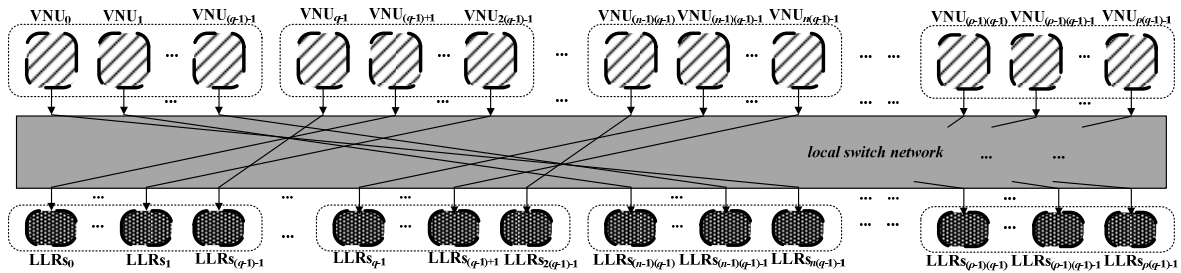

Figure 4: Local shuffle network of Class-I codes case.

TABLE I  COMPLEXITY COMPARISONS FOR DIFFERENT NON-BINARY QC-LDPC CODE DECODER NETWORKS

| Different designs | | Proposed designs | | | | References | |
|---|---|---|---|---|---|---|---|
| | | #1 | #2 | #3 | #4 | [4][a] | [5][b] |
| Global shuffle network | Wires[c] | $b_q n_m(q-1)d_c$ | $b_q n_m(q-1)d_c$ | $b_q n_m(q-1)d_c$ | $b_q n_m(q-1)d_c$ | $3b_q n_m(q-1)d_c$ | $3b_q n_m(q-1)d_c$ |
| | De-MUX's | 0 | $(q-1)\rho$ | $(q-1)\rho$ | $(q-1)\rho$ | $3(q-1)\rho$ | $3(q-1)\rho$ |
| | LUT bits | 0 | $p(q-1)\rho$ | $p(q-1)\rho$ | $p(q-1)\rho$ | $p(q-1)[\rho+\gamma(\gamma-1)/2]$ | $p(q-1)(3\rho+\gamma-2)$ |
| Local shuffle network | Wires | $b_q(q-1)\gamma$ | $b_q(q-1)\gamma$ | $b_q(q-1)\gamma$ | $2b_q(q-1)\gamma$ | – | – |
| | De-MUX's | 0 | 0 | 0 | $b_q(q-1)\gamma$ | – | – |
| | Crossbars | 0 | 0 | $\rho(\log_2\rho - 1/2)$ | $\rho(\log_2\rho - 1/2)$ | – | – |
| | LUT bits | 0 | 0 | $(\gamma\rho\log_2\rho)/2$ | $(\gamma\rho\log_2\rho)/2$ | – | – |
| Class-I code | | Yes | Yes | No | Yes | Yes | Yes |
| Class-II code | | No | No | Yes | Yes | No | No |
| Flexibility | | No | Yes | Yes | Yes | No | No |

[a,b]Both designs are for a 32-ary (837, 726) rate-0.85 Class-I code. [c]$n_m$ is the selecting parameter for Min-Max decoding algorithm.

## B. Local Shuffle Network for Class-II Codes

### 1) Generating Algorithm for Local Shuffle Network

Similar to Class-I codes, *local shuffle network* for Class-II codes can be constructed based on the symmetry properties:

**Scheduling Algorithm for Local Shuffle Network - II**

**1:** for all $0 < v \leq l$, the beginning of decoding the $v$th layer **do**
**2:**   for all $0 \leq i < \rho$ **do**
**3:**     for all $0 \leq j < q-1$ **do**
**4:**       Pass result of $\text{VNU}_{(q-1)(\textbf{index}_{v-1,i\bmod n}+n\lfloor i/n \rfloor)+j}$
**5:**       to $\text{VNU}_{(q-1)(\textbf{index}_{v,i\bmod n}+n\lfloor i/n \rfloor)+j}$
**6:**     endfor
**7:**   endfor
**8:** endfor

**INDEX**$^{(n)}$ is an $n \times n$ matrix of $[\textbf{index}_{i,j}]_{0 \leq i < n, 0 \leq j < n}$. Entries of the first row are defined by $\textbf{index}_{0,j} = j$. Other entries are derived from the *index assignment of surjective function* and symmetry properties. For instance, **INDEX**$^{(4)}$ is given by,

$$\textbf{INDEX}^{(4)} = \begin{bmatrix} 0 & 1 & 2 & 3 \\ 1 & 0 & 3 & 2 \\ 2 & 3 & 0 & 1 \\ 3 & 2 & 1 & 0 \end{bmatrix}. \quad (12)$$

A message shuffling example of 4-ary (12, 6) rate-½ Class-II code is illustrated in Fig. 5, the parameters are obtained by choosing $t = 1$, then $c = 2^{m-t} = 2$, and $n = 2^t = 2$ over $\textbf{GF}(2^2)$.

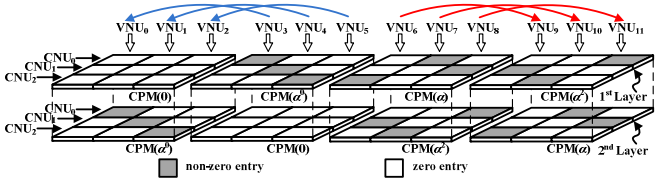

Figure 5: Layered decoding example of the 4-ary (12, 6) rate-½ Class-II code.

### 2) Achitecture of Local Shuffle Network

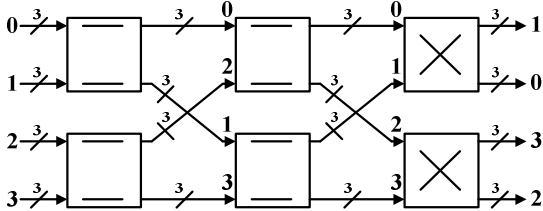

Figure 6: Local shuffle network of the Class-II code defined by Eq. (12).

The complexity of the resulting network is $1/q$ of that for the conventional one. Only $2\log_2\rho - 1$ stages and $\rho(\log_2\rho - 1/2)$ $2 \times 2$ crossbar switches are required. The number of control bits, which can be pre-acquired with **INDEX**$^{(n)}$ is $\rho(\log_2\rho - 1/2)$. The *local shuffle network* for Eq. (12) is given in Fig. 6. With a modified **INDEX**$^{(n)}$, this approach is suitable for Class-I codes.

## VI. IMPLEMENTATION COMPLEXITY COMPARISONS

Without loss of generality, it is assumed that the proposed $(u, v)$ decoder employs *Layer-I* partition. A $(b_q, b_f)$ uniform quantization is adopted, in which $b_f$ out of $b_q$ bits are used for fraction parts. Table I lists the comparison of the proposed design with others. Compared with the state-of-the-art decoders, the proposed one can greatly reduce the hardware complexity. According to [4] and [5], their network could not incorporate flexibility into the design due to the use of ROM. In Table I, there are two approaches to design the *local shuffle network* for Class-I codes. The first one (#1) is reconfigurable for any Class-I codes with code length $\rho(q-1)$. The second one (#2) is only suitable for a specific Class-I code. The #3 and #4 approaches deal with Class-II codes and codes of both classes, respectively. Take the 64-ary (1260, 630) rate-0.5 Class-I code as an example. While having more flexibility, the proposed shuffle network #1 achieves hardware saving of 69.2% and 70.6% compared with [4] and [5], respectively. Because the #2 approach can further eliminate all memory elements required by #1, more reduction can be expected. For configurable version of the 32-ary (992, 496) rate-0.5 Class-II code decoder with #3 network, the total saving is $(k-1)/k = 15/16 \approx 93.8\%$.

## VII. CONCLUSIONS

In this paper, with the exploited geometry properties of non-binary QC-LDPC codes, novel approaches to design efficient network for decoders are proposed, which outperform the state-of-the-art designs with more than 69.2% savings.